\documentclass[12pt]{article}
\usepackage{amsthm,amsfonts,amsmath,amssymb,mathrsfs,url}
\usepackage[usenames,dvipsnames]{color}

\title{Multi-Time Equations, Classical and Quantum}

\author{
S\"oren Petrat\footnote{Mathematisches Institut, 
	Ludwig-Maximilians-Universit\"at, Theresienstr. 39, 80333 M\"unchen, Germany.
	E-mail: petrat@math.lmu.de}\ \ and
Roderich Tumulka\footnote{Department of Mathematics,
     Rutgers University,
     110 Frelinghuysen Road, Piscataway, NJ 08854-8019, USA.
     E-mail: tumulka@math.rutgers.edu}
}
\date{January 9, 2014}

\theoremstyle{plain}
\theoremstyle{plain}
\theoremstyle{plain}
\theoremstyle{plain}
\theoremstyle{definition}

\newcommand{\be}{\begin{equation}}
\newcommand{\ee}{\end{equation}}
\newcommand{\Hilbert}{\mathscr{H}}

\newcommand{\RRR}{\mathbb{R}}
\newcommand{\CCC}{\mathbb{C}}

\newcommand{\vu}{\boldsymbol{u}}
\newcommand{\vv}{\boldsymbol{v}}
\newcommand{\vw}{\boldsymbol{w}}
\newcommand{\vp}{\boldsymbol{p}}

\newcommand{\vx}{\boldsymbol{x}}

\newcommand{\sS}{\mathscr{S}}

\newcommand{\foliation}{\mathscr{F}}

\begin{document}
\maketitle

\begin{abstract}
Multi-time equations are evolution equations involving several time variables, one for each particle. Such equations have been considered for the purpose of making theories manifestly Lorentz invariant. We compare their status and significance in classical and quantum physics.

\medskip

\end{abstract}

Motivated by relativity, several authors have suggested introducing, in classical or quantum physics, a separate time coordinate for every particle in order to avoid the use of simultaneity surfaces; see, e.g., \cite{dirac:1932,dfp:1932,bloch:1934,DV82a,LLP89,Dup13,pt:2013a,pt:2013c}. For example, multi-time wave functions have been used in the foundations of quantum mechanics for defining relativistic collapse \cite{Tum04} or de Broglie--Bohm \cite{HBD} theories.

In quantum physics, the basic idea is that the relativistic analog of a non-relativistic wave function
\be
\psi(t,\vx_1,\ldots,\vx_n)
\ee 
(say, a function on $\RRR\times(\RRR^3)^n$) is a multi-time wave function 
\be
\phi(t_1,\vx_1,\ldots,t_n,\vx_n)
\ee
defined, say, on $(\RRR^4)^n$ or on the set of spacelike configurations,
\be
\sS_n= \Bigl\{ (x_1,\ldots,x_n)\in(\RRR^4)^n: x_j \text{ is spacelike to } x_k \forall j\neq k \Bigr\}\,,
\ee
where we write $x=(x^0,\vx)=(t,\vx)$ for a space-time point. The ordinary wave function $\psi$ is retrieved by setting all time coordinates equal (in the Lorentz frame that $\psi$ refers to),
\be
\psi(t,\vx_1,\ldots,\vx_n) = \phi(t,\vx_1,\ldots,t,\vx_n)\,.
\ee
The time evolution of $\phi$ is governed by one Schr\"odinger equation per time coordinate,
\be\label{phiHj}
i\frac{\partial \phi}{\partial t_j} = H_j \phi
\ee
for all $j=1,\ldots,n$, with partial Hamiltonians $H_j$ that (at least at $t_1=t_2=\ldots = t_n$) add up to the full Hamiltonian 
\be
H=\sum_{j=1}^n H_j
\ee
that figures in the Schr\"odinger equation for $\psi$,
\be\label{psiH}
i\frac{\partial \psi}{\partial t} = H\psi\,.
\ee
The system of equations \eqref{phiHj} is consistent (i.e., integrable) only if the partial Hamiltonians satisfy the consistency condition \cite{bloch:1934,GU12,pt:2013a}
\be\label{consistency}
\frac{\partial H_k}{\partial t_j}- \frac{\partial H_j}{\partial t_k}+i \Bigl[ H_j,H_k \Bigr] =0\,. 
\ee
We can drop the assumption of a fixed particle number $n$ by using Fock space as the Hilbert space, $\Hilbert=\oplus_{n=0}^\infty \Hilbert_n$ with $\Hilbert_n$ the space of $n$-particle states. A vector $\psi_0$ in Fock space can be regarded, in the particle--position representation, as a function on $\Gamma(\RRR^3)$ with
\be
\Gamma(S):= \bigcup_{n=0}^\infty S^n
\ee
(the configuration space of a variable number of particles), or, equivalently, as a sequence $(\psi_0^{(0)}, \psi_0^{(1)}, \psi_0^{(2)},\ldots)$, where $\psi_0^{(n)}$, the $n$-particle sector of $\psi_0$, is a function on $(\RRR^{3})^n$. In this setting, the single-time wave function $\psi$ is a function on $\RRR\times \Gamma(\RRR^3)$, and its multi-time version $\phi$ is a function on $\Gamma(\RRR^4)$ or on the set of spacelike configurations
\be
\sS=\bigcup_{n=0}^\infty \sS_n\,.
\ee

Now consider again a fixed $n>1$. The equation in classical mechanics that is perhaps the closest analog of the Schr\"odinger equation \eqref{psiH} is the Hamilton--Jacobi equation
\be\label{SH}
\frac{\partial S}{\partial t}(t,\vx_1,\ldots,\vx_n) 
= -H\bigl(t,\vx_1,\nabla_{\vx_1} S,\ldots,\vx_n, \nabla_{\vx_n} S\bigr)
\ee
with classical Hamiltonian function $H(t,\vx_1,\vp_1,\ldots,\vx_n,\vp_n)$, and the closest analog of the wave function $\psi$ of quantum mechanics is perhaps the Hamilton--Jacobi function $S(t,\vx_1,\ldots,\vx_n)$. In this spirit, the analog of the multi-time wave function $\phi$ would be a multi-time Hamilton--Jacobi function $S(t_1,\vx_1,\ldots,t_n,\vx_n)$ satisfying a system of multi-time Hamilton--Jacobi equations (e.g., \cite[Eq.~(3.59)]{LLP89}, \cite{UMD09}, \cite[Eq.~(48)]{Dup13})
\be\label{SHj}
\frac{\partial S}{\partial t_j}(t_1,\vx_1,\ldots,t_n,\vx_n) 
= -H_j\bigl(t_1,\vx_1,\nabla_{\vx_1} S, \ldots,t_n,\vx_n, \nabla_{\vx_n} S\bigr)
\ee
with partial Hamiltonian functions $H_j(t_1,\vx_1,\vp_1,\ldots,t_n,\vx_n,\vp_n)$ that (at least at $t_1=t_2=\ldots=t_n$) add up to $H$. The system of equations \eqref{SHj} is consistent only if the partial Hamiltonians satisfy the consistency condition (e.g., \cite[Eq.~(3.60)]{LLP89}, \cite[Eq.~(50)]{Dup13}, \cite[Eq.~(2.2)]{GU12})
\be\label{Sconsistency}
\frac{\partial H_k}{\partial t_j}- \frac{\partial H_j}{\partial t_k}- \Bigl\{ H_j,H_k \Bigr\} =0\,,
\ee
where $\{\cdot,\cdot\}$ denotes the Poisson bracket
\be
\{f,g\} = \sum_{j=1}^n \nabla_{\vx_j}f \cdot \nabla_{\vp_j} g - \nabla_{\vp_j} f \cdot \nabla_{\vx_j} g\,.
\ee
The analogy between \eqref{consistency} and \eqref{Sconsistency} is striking, in particular since it is widely regarded as a quantization rule to replace the Poisson bracket by $-i$ times the commutator.

However, the analogy is limited. The main difference is that for a quantum system with a wave function, there is a fact in nature about what its wave function is, while in classical mechanics there is no fact in nature about what its Hamilton--Jacobi function $S$ is. In classical mechanics, nature need only know the present positions and momenta of all particles to determine the future evolution; thus, of $S$ nature need only know $\nabla_{\vx_j}S$ for all $j$ at the actual configuration, while all other information about $S$ is irrelevant to the actual evolution of the system. In short, $\psi$ is real but $S$ is not.\footnote{Some readers may not agree that nature knows what $\psi$ is. To them, the Schr\"odinger equation will seem less fundamental, its manifest covariance less relevant, and multi-time equations less attractive.} As a consequence, the Hamilton--Jacobi equation \eqref{SH} can hardly be regarded as a fundamental law of nature in classical mechanics. The fundamental law would be the equations that directly govern the motion, which can be written as Hamilton's equations
\begin{subequations}\label{Ham}
\begin{align}\label{Ham1}
\frac{d\vx_j}{dt} &= \:\:\:\nabla_{\vp_j} H\Bigl(t,\vx_1(t),\vp_1(t),\ldots,\vx_n(t),\vp_n(t)  \Bigr)\,,\\
\label{Ham2}\frac{d\vp_j}{dt} &= -\nabla_{\vx_j} H\Bigl(t,\vx_1(t),\vp_1(t),\ldots,\vx_n(t),\vp_n(t) \Bigr)\,.
\end{align}
\end{subequations}
Thus, in a relativistic formulation of classical mechanics, it would seem more appropriate to look for a multi-time version of \eqref{Ham} than of \eqref{SH}. It will be useful to write the equations of motion in an even more abstract way, as
\begin{subequations}\label{vw}
\begin{align}\label{vw1}
\frac{d\vx_j}{dt} &= \,\vv_j\Bigl(t,\vx_1(t),\vp_1(t),\ldots,\vx_n(t),\vp_n(t) \Bigr)\,,\\
\label{vw2} \frac{d\vp_j}{dt} &= \vw_j\Bigl(t,\vx_1(t),\vp_1(t),\ldots,\vx_n(t),\vp_n(t)  \Bigr)\,,
\end{align}
\end{subequations}
where $(\vv_1,\vw_1,\ldots,\vv_n,\vw_n)$ is a (time-dependent) vector field on phase space that governs the motion. The obvious multi-time version of \eqref{vw} reads
\begin{subequations}\label{vwj}
\begin{align}\label{vw3}
\frac{d\vx_j}{dt_j} &=\, \vv_j\Bigl(t_1,\vx_1(t_1),\vp_1(t_1),\ldots,t_n,\vx_n(t_n),\vp_n(t_n) \Bigr)\,,\\
\label{vw4} \frac{d\vp_j}{dt_j} &= \vw_j\Bigl(t_1,\vx_1(t_1),\vp_1(t_1),\ldots,t_n,\vx_n(t_n),\vp_n(t_n)  \Bigr)\,,
\end{align}
\end{subequations}
and the relevant consistency property means the following. Consider the actual history of the classical system, consisting of $n$ timelike particle world lines; we will call this an $n$-path. For each $j=1,\ldots,n$ choose a point $x_j=(t_j,\vx_j)$ on the $j$-th world line so that $(x_1,\ldots,x_n)$ is a spacelike configuration. Let $\vp_j$ be the corresponding momentum, $\vp_j=m_j \vu_j$ with $u_j=(u_j^0,\vu_j)$ be the future-pointing unit tangent vector to the $j$-th world line at $x_j$. As we vary $t_j$, $\vx_j$ and $\vp_j$ will vary. We regard \eqref{vwj} as \emph{valid} multi-time equations if and only if  they correctly represent how $\vx_j$ and $\vp_j$ vary with $t_j$, for any spacelike choice of $(x_1,\ldots,x_n)$ on the $n$ world lines. Furthermore, we regard \eqref{vwj} as a \emph{consistent} system if and only if for (almost) every spacelike configuration $(x_1,\ldots,x_n)$ there is an $n$-path passing through $(x_1,\ldots,x_n)$ for which \eqref{vwj} is valid. Then the necessary and sufficient condition for consistency (e.g., \cite[Eq.~(1.2)]{DV82a}, \cite[Eq.~(2.7)]{LLP89}) is that, for all $j\neq k$,
\begin{subequations}\label{dconsistency}
\begin{align}
\biggl( \frac{\partial}{\partial t_j} + \vv_j \cdot \nabla_{\vx_j} + \vw_j \cdot \nabla_{\vp_j} \biggr) \vv_k\, 
&=0 ~~\text{and}\\ 
\biggl( \frac{\partial}{\partial t_j} + \vv_j \cdot \nabla_{\vx_j} + \vw_j \cdot \nabla_{\vp_j} \biggr) \vw_k &=0\,, 
\end{align}
\end{subequations}
a condition rather different from \eqref{consistency}, and a condition that does not follow from \eqref{Sconsistency}.\footnote{We note that, remarkably, \eqref{Sconsistency} is also equivalent to the consistency (i.e., integrability) (e.g., \cite[Eq.~(2.10)]{LLP89}, \cite[Eq.~(55)]{Dup13}, \cite[Eq.~(2.2)]{GU12}) of the following system of equations (e.g., \cite[Eq.~(2.12)]{LLP89}, \cite[Eq.~(53)--(54)]{Dup13}), writing $\vec{t}=(t_1,\ldots,t_n)$:
\begin{subequations}\label{xt1tn}
\begin{align}
\frac{\partial}{\partial t_k}\vx_j\bigl(\vec{t}\,\bigr) 
&= \:\:\:\nabla_{\vp_j} H_k\Bigl(t_1,\vx_1\bigl(\vec{t}\,\bigr),\vp_1\bigl(\vec{t}\,\bigr),\ldots,\vx_n\bigl(\vec{t}\,\bigr),\vp_n\bigl(\vec{t}\,\bigr)  \Bigr)\,,\\
\frac{\partial}{\partial t_k} \vp_j\bigl(\vec{t}\,\bigr)
&= -\nabla_{\vx_j} H_k\Bigl(t_1,\vx_1\bigl(\vec{t}\,\bigr),\vp_1\bigl(\vec{t}\,\bigr),\ldots,\vx_n\bigl(\vec{t}\,\bigr),\vp_n\bigl(\vec{t}\,\bigr)  \Bigr)\,.
\end{align}
\end{subequations}
This system can be regarded as another multi-time variant of \eqref{Ham}. However, a function of the form $\vx_j\bigl(\vec{t}\,\bigr)$ typically does not define a world line (while one of the form $\vx_j(t_j)$ does).}

By the last remark we mean the following: Even if \eqref{Sconsistency} is satisfied, and \eqref{SHj} can be solved, then the resulting multi-time Hamilton--Jacobi function $S(t_1,\vx_1,\ldots,t_n,\vx_n)$ does not necessarily provide us with a $3n$-parameter family of $n$-paths, as the equations of motion, (e.g.)\footnote{Equations \eqref{dvx_jdt_jS} and \eqref{dvx_jdtS} are actually non-relativistic, but that does not matter for illustrating the point we are making here, which concerns the multi-time consistency condition corresponding to \eqref{dconsistency}, a condition that can also be considered in the non-relativistic case. The relativistic formulas would only introduce further complications.} 
\be\label{dvx_jdt_jS}
\frac{d\vx_j}{dt_j} = \frac{1}{m_j} \nabla_{\vx_j} S(t_1,\vx_1,\ldots,t_n,\vx_n)\,,
\ee
which form the natural analog of the equation of motion of the standard single-time Hamilton--Jacobi formalism, (e.g.)
\be\label{dvx_jdtS}
\frac{d\vx_j}{dt} = \frac{1}{m_j} \nabla_{\vx_j} S(t,\vx_1,\ldots,\vx_n)
\ee
are typically inconsistent in the sense that there is no $n$-path such that \eqref{dvx_jdt_jS} holds at every spacelike configuration $(x_1,\ldots,x_n)$ with $x_j$ lying on the $j$-th world line. Indeed, the condition for consistency of \eqref{dvx_jdt_jS} in this sense reads 
\be\label{Sdconsistency}
\biggl( \frac{\partial}{\partial t_j} + \frac{1}{m_j} \nabla_{\vx_j} S(t_1,\vx_1,\ldots,t_n,\vx_n) \cdot \nabla_{\vx_j} \biggr) \nabla_{\vx_k} S(t_1,\vx_1,\ldots,t_n,\vx_n)=0\,,
\ee
and this is typically not fulfilled (except in the non-interacting case). Put differently, if we choose a spacelike foliation $\foliation$ and demand \eqref{dvx_jdt_jS} to hold only on configurations that are simultaneous with respect to $\foliation$, then we obtain an $n$-path for any initial configuration, but then different choices of $\foliation$ will typically lead to different $3n$-parameter families of $n$-paths for the same $S(t_1,\vx_1,\ldots,t_n,\vx_n)$ function. This kind of situation is well known in Bohmian mechanics, see, e.g., \cite{Tum07}: Even if the multi-time equations are consistent and the multi-time wave function $\phi$ is well defined, then the Bohmian equation of motion, analogous to \eqref{dvx_jdt_jS}, usually yields a different set of particle world lines for every foliation $\foliation$. Thus, the meaning of \eqref{dconsistency} is neither a direct analog of \eqref{Sconsistency} nor of \eqref{consistency}.

The disanalogy goes further. We are not aware of any relativistic, classical, physical theory of interacting particles that can be formulated in the form \eqref{vwj}. Relativistic interaction is usually mediated by fields, and action-at-a-distance (say, via interaction potentials) perhaps should not be expected to be relativistically invariant. In fact, Currie, Jordan, and Sudarshan \cite{CJS63} have shown for $N=2$ particles that, in a certain Hamiltonian framework, any law of motion \eqref{vwj} that satisfies \eqref{dconsistency} and is Lorentz invariant must be non-interacting (i.e., the particle world lines are straight lines).\footnote{The framework requires that laws of motion be of Hamiltonian form, with the Hamiltonian part of a representation of the Poincar\'e group on phase space by transformations that preserve the symplectic structure. We also note that their result is actually slightly stronger than what we just described, requiring only the slightly weaker assumption that the law of motion be valid on configurations that lie on a common spacelike hyperplane.} Therefore, multi-time equations of the form \eqref{vwj} may not be a viable formulation of classical relativistic theories. 

The situation is rather different in quantum physics, where multi-time Schr\"odinger equations do seem like a possible, even natural, formulation \cite{dfp:1932,bloch:1934,pt:2013c,pt:2013d}. This is mainly because the relevant kind of interaction is based on particle creation and annihilation, which does not fit into the framework of classical mechanics in terms of ordinary differential equations such as \eqref{Ham} but fits very well into Schr\"odinger equations for vectors in Fock space. In fact, since interaction potentials (given by multiplication operators) lead to violation of the consistency condition \eqref{consistency} also in the quantum case \cite{pt:2013a}, it would seem that without particle creation and annihilation we might be stuck with non-interacting particles. A multi-time formulation of quantum theory with particle creation can be set up as follows \cite{pt:2013c,pt:2013d}. Since a vector in Fock space $\oplus_{n=0}^\infty S_{\pm} L^2(\RRR^3,\CCC^k)^{\otimes n}$ (with $S_{\pm}$ the (anti-)symmetrizer) can be regarded as a function $\psi$ on $\Gamma(\RRR^3)$, its multi-time analog is a function $\phi$ on $\Gamma(\RRR^4)$, or rather on the set $\sS$ of spacelike configurations. Its multi-time evolution is governed by $n$ equations for its $n$-particle sector, equations that may also involve (e.g.) the $n-1$ and the $n+1$-particle sector of $\phi$. The consistency question is then more delicate but can be answered, in fact positively for natural examples of multi-time equations \cite{pt:2013c,pt:2013d}.

As a last remark, as soon as fields are introduced in classical relativistic physics for mediating the interaction, there is apparently no need any more for multiple time variables. For example, the equations of classical electrodynamics of $n$ charged particles are\footnote{The system of equations \eqref{CED} is actually ultraviolet divergent and therefore ill defined, but that is an issue orthogonal to the ones considered in this paper.}
\begin{subequations}\label{CED}
\begin{align}
m_j \frac{d^2x^\mu_j}{d\tau^2} &= q_j  \, F^\mu_{\:\:\:\:\nu}(x_j(\tau))\, \frac{dx^\nu_j}{d\tau} ~~~~~ \forall j\in\{1\ldots n\}\,,\label{Lorentz}\\
\partial_\mu F^{\mu\nu}(x)&= 4\pi\sum_{j=1}^n q_j\, \int d\tau\, \delta^4\bigl(x-x_j(\tau)\bigr) \, \frac{dx^\nu_j}{d\tau}\,, \label{Max1}\\
\partial_{\lambda}F_{\mu\nu} + \partial_\mu F_{\nu\lambda} + \partial_\nu F_{\lambda\mu}&=0\,, \label{Max2}
\end{align}
\end{subequations}
where \eqref{Lorentz} is the equation of motion including the Lorentz force law, $\tau$ means proper time along the world line, the constant $q_j$ is the charge of particle $j$, $F_{\mu\nu}$ is the electromagnetic field tensor (a function on space-time $\RRR^4$), and \eqref{Max1}--\eqref{Max2} are the Maxwell equations. These equations are manifestly Lorentz invariant and obviously do not require mentioning any function of several time variables. The reason multiple times are not needed is that the motion of particle $j$ at $x_j$ does not depend on the other particles except through the field at $x_j$, and thus has to do with the fact that classical electrodynamics is local. In contrast, quantum physics is notoriously non-local, and this is reflected in the dependence of the wave function on the positions of several particles---which is the reason for introducing several time variables.

\bigskip

\noindent{\it Acknowledgments.} We thank Steven Duplij and Matthias Lienert for helpful discussions. S.P.\ acknowledges support from Cusanuswerk, from the German--American Fulbright Commission, and from the European Cooperation in Science and Technology (COST action MP1006). R.T.\ acknowledges support from the John Templeton Foundation (grant no.\ 37433) and from the Trustees Research Fellowship Program at Rutgers. 

\end{document}